\documentstyle [12pt]{article}

\def\be{\begin{equation}}
\def\ee{\end{equation}}
\def\bea{\begin{eqnarray}}
\def\eea{\end{eqnarray}}

\begin{document}

\begin{titlepage}

\title{Quantized spaces are four-dimensional compact manifolds with de-Sitter
($O(1,4)$ or $O(2,3)$) group of motion}
\author{A.N. Leznov\\
Universidad Autonoma del Estado de Morelos,\\ 
CIICAp,Cuernavaca, Mexico}

\maketitle

\begin{abstract}

It is shown uniquelly that quatized spaces are realised on four-dimensional
compact manifolds. In the case of  $O(1,5)$ quntaized space this are four independent
parameters of $O(5)$ unite vector; in the case of $O(2,4)$ these are
parameters of one two-dimensional unite vector ( 1 parameter) and components of unite four-
dimensional vector (3- parameters) and at last in the case $O(3,3)$ these are parameters of 
2 independent 3-dimensional unite vectors (each have 2 parameters). This result 
follows directly only from the condition to have a correct limit to ussual theory 
(correspondence principle).
\end{abstract}

\end{titlepage}


\section{Introduction}

In the previous paper of the author \cite{I} it was lost and not used very important condition which allow to reconstruct uniquelly
representation of six-dimensional rotation algebra of quantized space. It was not taking in acount the fact that that generators of 
Lorenz rotations in the limit to the ussual theory ($L^2\to \infty,M^2\to \infty,H\to \infty$) are algebraically connected with the 
generators of coordnates and impulses by the well known quadratical relation:
\be
S_{i,j}\equiv F_{i,j}+p_ix_j-p_jx_i=0\label{KL}
\ee
As a consequence two Kazimir operators of Lorenze group ( generated by $S_{ij}$) $K_2=\sum S_{i,j}S_{i,j}, 
K_3=\sum \epsilon_{i,j,k,l} S_{i,j}S_{k,l}$ are also equal to zero. But inverse is not true. From the fact that two Kazimir generators 
of Lorenze group (it is not compact (!)) are equal to zero does't follow that representation is the trivial one. Directly from
(\ref{KL}) obvious additional relations follows $\tilde x_i\equiv \epsilon_{i,j,k,l} x_j F_{k,l}=0,
\tilde p_i\equiv \epsilon_{i,j,k,l} p_j F_{k,l}=0, \epsilon_{i,j,k,l} F_{i,j} F_{k,l}=0$, which will be important for consideration
below.  

In the the previous paper \cite{I} it was assumed that representation of the algebra of quantized
space must be choosen in such way that in the ``classical limit'' ($L^2\to \infty,M^2\to \infty,H\to \infty$) second order Kazimir 
operator passes to unity and both other ones pass to zero. But consideration above shows that really in classical limit must be 
satisfied more strong condition  (\ref{KL}). Compare the similar consideration in \cite{LMOD} and \cite{LMOST}. Representation of 
this kind was called their as scalar one.

Fortunelly as it will be shown below such representations are existed. More other these representations are realised on 
four-dimensional space of compact parameters. Thus quantized spases are also four-dimensional but compact manifolds in comparision 
with noncompact space-time manifold of the ussual theory.

A little additional explanation. Six-dimensional rotation groups of the theory of quatized space 
are 15-th parametrical. They posses 3 Kazimir operators. Thus in the general case their irreducible 
representation (of the general position with three independent Kazimir operators) may be realized on the space of 
${15-3\over 2}=6$ parameters. Exactly on this number of parameters was realized the algebras of the real forms 
$O(3,3),O(2,4).O(1,5)$ in \cite{I}. But the boundary condition in \cite{I} was choosen not correctly. The correct choice 
(\ref{KL}) leads to four-dimensional compact quantized spaces.

The goal of the present paper is to explain and clarify this situation.

In what follows we preserve all notations of \cite{I}.

\section{Kazimir operators and general stratedgy}

For convinience of the reader we rewrite commutation relations of quantized space from \cite{I},\cite{1}
We will work below with the algebra of quantum space proposed in \cite{1} and containing 3 dimensional
parameters of the square of the length $L^2$, square of impulse $M^2$ and action
$H$. Commutation relations of such algebra have the following form
$$
[p_i,x_j]=ih(g_{ij}I+{F_{ij}\over H}),\quad  [p_i,p_j]={ih\over L^2}F_{ij},
\quad [x_i,x_j]={ih\over M^2}F_{ij},
$$
\be
[I,p_i]=ih({p_i\over H}-{x_i\over L^2}), \quad [I,x_i]=ih({p_i\over M^2}-
{x_i\over H}),\quad [I,F_{ij}]=0 \label{QS}
\ee
$$
[F_{ij},x_s]=ih(g_{js}x_i-g_{is}x_j),\quad [F_{ij},p_s]=ih(g_{js}p_i-g_{is}p_j)
$$
$$
[F_{ij},F_{sk}]=ih(g_{js}F_{ik}-g_{is}F_{jk}-g_{jk}F_{is}+g_{ik}F_{js})
$$
 
We present the explicit expressions for Kazimir operators from \cite{I}
$$
K_2=-I^2+{1\over L^2}p^2-{1\over M^2}p^2-{(px)+(xp)\over H}+
({1\over H^2}-{1\over L^2M^2})(l^2-f^2)=
$$
$$
\nu^2(-{I^2\over \nu^2}+{(x-{L^2\over H}p)^2\over \nu^2 L^2}+{L^2\over H^2}p^2+
{(f^2-l^2)\over H^2})
$$
where  $a^2=a_1^2+a_2^2+a_3^2-a_4^2,l^2=(\vec l)^2,f^2=(\vec f)^2$ squares of three 
dimensional generators of rotations and Lorenz boosts. In what follows $\bar x\equiv (x-{L^2\over H}p)$.
Generators of three-dimensional space rotation are a compact ones and thus all generators (their squares) with 
the same sign in $K_2$ are also compact, with the opposit are noncompact. In the case 
$0 \leq L^2,M^2,\nu^2={H^2\over L^2M^2}-1$ this is $O(3,3)$ algebra and so on.  

Two other operators of Kazimir in three and four dimensional notations are as follows
$$
K_3=I(\vec f \vec l)+(\vec f [\vec p,(\vec x-{L^2\over H}\vec p))-(\vec l, (p_4
(\vec x-{L^2\over H}\vec p)-\vec p(x_4-{L^2\over H}p_4))
$$ 
$$
K_4=\sum_{i\leq j} S_{i,j}S_{i,j} -{1\over L^2}(\tilde x_i)^2+{1\over H}(\tilde x_i\tilde p_i+
\tilde p_i\tilde x_i)-{1\over M^2}(\tilde p_i)^2+({1\over L^2M^2}-{1\over H^2})(f,l)^2
$$
where generators of the "spin" variables is defined as $S_{i,j}\equiv IF_{i,j}+p_ix_j-p_jx_i$ and 
four-dimensional vectors $\tilde x_i,\tilde p_i$ ( of the psevdo coordinates and psevdo impulses) are 
defined as $\tilde x_i=\sum \epsilon_{ijkl} x_jF_{kl},\tilde p_i=\sum \epsilon_{ijkl}p_jF_{kl}$. 
Psevdovectors $\tilde{\bar x_i},\tilde p_i$ are introduced in analogie of 
Pauli-Lubansky  vector in representation theory of the Poincare algebra. In the classical limit 
the operators of Cazimir $K_3,K_4$ fixed quantum numbers of representation of the Lorentz $O(1,3)$ algebra.

[Without any connection with what follows we pay attention of the reader on 
symmetry of Kazimir operators of the second and fourth order with respect to substitution $\vec l,\vec f
\to S,I\to (\vec f\vec l),x\to \tilde x, p\to \tilde p$.]

As was mentioned in introduction the classical limit  would be satisfied under the assumtion that 
in representation of the quantum space algebra $S_{i,j}=IF_{i,j}+p_i\bar x_j-p_j\bar x_i=0$. 
In the classical limit $I\to 1$ and condition (\ref{KL}) would be a direct consequent  
of the last choice of the representation of six-dimensional group of rotation. To have the zero value 
for the Kazimir operators of 3 and 4 order it is sufficient to assume additionally
$$
\tilde {\bar x_i}=0,\quad \tilde p_i=0,\quad (lf)=0
$$
Thus if it will be possible to satisfy 15 conditions above then we will have representation of 
six-dimensional algebra having correct limit to 
four dimensional coordinate space of the ussual theory. Below we rewrite these equation in three
-dimensional notations and after this will try to resolve them for all real forms of ortogonal group 
(find such representations in which they are satisfied):
$$
I f_{\alpha}+p_{\alpha} \bar x_4-p_4 \bar x_{\alpha}=0,\quad I L_{\alpha,\beta}+p_{\alpha}\bar x_{\beta}- 
p_{\beta}\bar x_{\alpha}=0
$$
\be
\bar x_4 L_{\alpha,\beta}+\bar x_{\alpha}f_{\beta}-\bar x_{\beta}f_{\alpha}=0,\quad
p_4 L_{\alpha,\beta}+p_{\alpha}f_{\beta}-p_{\beta}f_{\alpha}=0\label{BC}
\ee
$$
(fl)=(pl)=(\bar xl)=0
$$
Not all of this equations are independent but this is not essential for further 
consideration.

Now it is possible to say that that quantized space is described by commutation relations (\ref{QS}) and
additional algebraic equations (\ref{BC}), which responcible for the  correct limit to the ussual theory 
(correspondence principle) and which must be satisfied by the choice of corresponding representation of 
six-dimensional algebra of quantized space. 

\section{$O(3,3)$ case with $SO(2,3)$ group of motion}

$$
K_2=\nu^2({p^2L^2\over H^2}+{\bar x^2\over L^2\nu^2}+{f^2-l^2\over H^2}-{I^2\over \nu^2}),\quad 0\leq L^2, \nu^2=
{H^2\over L^2M^2}-1
$$
In this case 6 generators 
$$
I=F_{65},\quad p_4=F_{45},\quad \bar x_4=F_{46},\quad l_{\alpha}=
\epsilon_{\alpha,\beta,\gamma}Q_{\beta,\gamma}
$$ 
are compact one and 9
$$
p_{\alpha}=F_{\alpha,6},\quad \bar x_{\alpha}=F_{\alpha,5},\quad f_{\alpha}=F_{\alpha,4}
$$ 
are non compact.

In connection with \cite{I}   
\be
F_{\alpha,i}=\sum \rho_{\sigma} q^{\sigma}_{\alpha} p_i^{\sigma}+\sum_{\delta \leq j}
q^{\delta}_{\alpha} p_i^j P^{j,\delta}+\sum_{j\leq \delta}
q^{\delta}_{\alpha} p_i^j Q^{\delta,j}\label{BF}
\ee
Left and right shifts of compact $O(3)$ groups are connected by the relations
$$
Q_{\alpha,\beta}=\sum_{\mu,\nu} q_{\alpha}^{\nu}q_{\beta}^{\mu} Q^{\nu,\mu},\quad
P_{i,j}=\sum_{k,l} p_i^kp_j^l P^{k,l}
$$

\subsection{Reducibility of representation.Invariant subspaces}

Let us first consider 3 three scalar equations (\ref{BC}) $(fl)=(pl)=(\bar xl)=0$. 
Substituting all above expressions and taking into acount the obvious equality
$$
\sum_{\alpha,\beta,\gamma} q_{\alpha}^{\nu}q_{\beta}^{\mu}q_{\gamma}^{\delta}=
\epsilon_{\nu,\mu,\delta}
$$
we come to equation
$$
\sum \rho_{\sigma} \epsilon_{\sigma,\mu,\nu}p_i^{\sigma}Q^{\mu,\nu}+\sum_{\delta \leq j}
\epsilon_{\delta,\mu,\nu} p_i^j Q^{\mu,\nu}P^{j,\delta}+\sum_{j\leq \delta}
\epsilon_{\delta,\mu,\nu} p_i^j Q^{\mu,\nu}Q^{\delta,j}=0
$$ 
3 components of this system transforms to 3 conditions
$$
(\rho_1+1)Q^{23}=0,\quad \rho_2 Q^{31}+(P^{21}+Q^{21})Q^{23}=0,\quad
\rho_3 Q^{12}+P^{31}Q^{23}+Q^{31}P^{32}=0
$$
By the same technique 3 equations $I f_{\alpha}+p_{\alpha} \bar x_4-p_4 \bar x_{\alpha}=0$ lead to
$$
(\rho_1+1)P^{23}=0,\quad \rho_2 P^{31}+(P^{21}+Q^{21})P^{23}=0,\quad
\rho_3 P^{12}+P^{31}Q^{32}+Q^{31}P^{23}=0
$$
From the above results it follows that the representation  with $\rho_2=\rho_3=0$ is space reducible.
One of its irreducible components is realized on subspace defined by the conditions $P^{23}=Q^{23}=0$.
 
This representation is realised exactly on two three dimensional unites vectors $p^1_i,
q^1_{\alpha}$. Substituting this condition into (\ref{BF}) we obtain
\be 
F_{\alpha,i}=\rho q^1_{\alpha} p_i^1+q^1_{\alpha} \sum_jp_i^j P^{j,1}+p_i^1\sum_{\delta}
q^{\delta}_{\alpha} Q^{\delta,1}\label{VIC}
\ee

9 remaining equations follows from (\ref{BC}) in terms of (\ref{VIC}) may be rewrriten as follows
$$
P_{i,j} Q_{\alpha,\beta}+F_{i,\alpha}F_{j,\beta}-F_{i,\beta}F_{j,\alpha}=0
$$
Direct check show that they are satisfied for representation generating by (\ref{VIC}).

This is finally expression for representation of $O(3,3)$ which is realised on two unite 
3-dimensional vectors $q^1_{\alpha},p^1_i$ or on 4 compact parameters. Explicit expression for 
these generators in terms of 4 parameters of two unite 3-dimensional vectores see 
in subsection below.

\subsection{Unitary representations}

As it follows directly from (\ref{VIC}) noncompact generators are antihermitians under the choice
$\rho=-2+i\sigma$. To check this it is necessary to consider equation $(F_{\alpha,i})^H=-F_{\alpha,i}$, 
keeping in mind that all compact generators are antihermitians $(P^{j,i})^H=-P^{j,i},
(Q^{\alpha,\delta})^H=-Q^{\alpha,\delta}$ and taking into account commutation relation $[Q^{\delta,1},
q^{\delta}_{\alpha}]=-q^1_{\alpha}$. But as was mentioned in \cite{I} such unitary
representation is in contradiction with classical limit $I\to 1$. Fortunely representation (\ref{VIC}) 
posses discrete serie which gives oposite sign for second Kazimir operator and leads to correct limit 
to ussual theory.

To find this representation it is nesessary to consider corresponding Hermitian form defined as
$$
K=\int dp^1dq^1 F^*(p^1,q^1) (K(p^1,q^1;\bar p^1,\bar q^1)F(\bar p^1,\bar q^1) d\bar p^1d\bar q^1
$$ 
were $dp^1=\sin \theta d\theta d\phi,\quad dq^1=\sin \tau d\tau d\psi$ is the invariant mesure on 
$O(3)$ and $(K(p^1,q^1;\bar p^1,\bar q^1)$ is the kernal of hermitian form which have to defined 
from the condition of its invariance with respect to transformation of the considered representation.

Condition of invariance $K$ with respect to compact transformations resticted the form the kernel 
function up to dependence of two parameters $K=K((p^1\bar p^1)[=x],(q^1\bar q^1)[=y]$ ($(p^1\bar p^1)=
\sum_1^3 p^1_i\bar p^1_i$ and so on).

After not combersom calculations we obtain the following system of equation with respect to kernel function:
$$
[(\rho*+4)+(\rho+4)xy-y(1-x^2)\partial_x-x(1-y^2)\partial_y] K=0,
$$
$$   
[(\rho*+4)xy+(\rho+4)-y(1-x^2)\partial_x-x(1-y^2)\partial_y] K=0
$$
$$
[(\rho*+4)x+(\rho+4)y-(1-x^2)\partial_x-(1-y^2)\partial_y] K=0
$$
$$
[(\rho*+4)y+(\rho+4)x-(1-x^2)\partial_x-(1-y^2)\partial_y] K=0
$$
Reducing first-second and third-fourth equations leads to equalities
$$
(\rho*-\rho)(1-xy)=0,\quad (\rho*-\rho)(x-y)=0
$$
from which it follows $\rho*=\rho$ or $x=1,y=1$ or in other words $K=\delta (1-x)\delta (1-y)$ in the 
second case. The last possibility is equivalent to considered above unitary representation with  
$\rho=-2+i\sigma$. For the first one simple resolving of the sytem of equations leads to: 
$$
K=c|x-y|^{-(\rho+4)}
$$
In the case of positive natural $\rho$ the last expression have to be considered as a 
generalised function or distribution \cite{GSH} and as it follows from the general theory 
it arised some number of invariant spaces on some of which unitary representation is 
realized (it is necessary additional check that hermitian form is positive defined).

\subsection{Explicit expressions. P-representation}

In this subsection we would like to rewrite all expessions for the generators of quantized
space under considration in terms of only four compact angles $p^1_1=\cos \theta,
p^1_2=\sin \theta \cos \phi,p^1_3=\sin \theta \sin \phi),q^1_1=\cos \tau,q^1_2=
\sin \tau \cos \psi,q^1_3=\sin \tau \sin \psi)$. In this notations
$$
P_{12}=-\cos \phi \partial_{\theta}+\sin \phi \cot \theta \partial_{\phi},\quad
P_{13}=-\sin \phi \partial_{\theta}-\cos \phi \cot \theta \partial_{\phi},\quad P_{23}=
-\partial_{\phi}
$$
$$
Q_{12}=-\cos \psi \partial_{\tau}+\sin \psi \cot \tau \partial_{\psi},\quad
Q_{13}=-\sin \psi \partial_{\tau}-\cos \psi \cot \tau \partial_{\psi},\quad Q_{23}=
-\partial_{\psi}
$$
For arising in (\ref{VIC}) operators $\hat P_i\equiv \sum_j p_i^j P^{j,1}=\sum_j p_j^1 
P_{j,i}$
with the help of the formulae above we have
$$
\hat P_1=\sin \theta \partial_{\theta},\quad \hat P_2=-\cos \phi \cos \theta \partial_
{\theta}+{\sin \phi \over \sin \theta} \partial_{\phi},\quad
\hat P_3=-\sin \phi \cos \theta \partial_{\theta}-{\cos \phi \over \sin \theta} 
\partial_{\phi}
$$
$$
\hat Q_1=\sin \tau \partial_{\tau},\quad \hat Q_2=-\cos \psi \cos \tau \partial_{\tau}+
{\sin \psi \over \sin \tau} \partial_{\psi},\quad
\hat Q_3=-\sin \psi \cos \tau \partial_{\tau}-{\cos \psi \over \sin \tau} 
\partial_{\psi}
$$
We present also explicit expresion for d Alamber operator- Kazimir operator of the 
second order of de-Sitter $O(2,3)$ algebra. In calculations below $p_6^1=p_1^1,p_5^1=
p_2^1,p_4^1=p_3^1$.    
$$
m^2=p^2+f^2-p_4^2-l^2=\sum [q^1_{\alpha}(\rho p_4^1+P_4)+p_4^1 Q_{\alpha}]^2+
\sum [q^1_{\alpha}(\rho p_5^1+P_5)+p_5^1 Q_{\alpha}]^2-p_4^2-l^2=
$$
(in formulae above $P_i=\sum_j p_i^j P^{j,1},Q_{\alpha}=\sum_{\delta}q^{\delta}_{\alpha}
Q^{\delta,1})$
$$
((\rho+1) p_4^1+P_4)^2+((\rho+1) p_5^1+P_5)^2-p_4^2-(p_6^1)^2l^2-(1+(p_6^1)^2)=
$$
$$
(\rho+1)(\rho+3)+P^2_3-[(\rho+1)\cos \theta- \sin \theta \partial_{\theta}]^2-p_4^2-
(p_6^1)^2l^2-(1+(p_6^1)^2)
$$
where now $P^2_3=\partial^2_{\theta}+\cot \theta \partial_{\theta}+{1\over \sin^2 
\theta}\partial^2_{\phi}$ Kazimir operator of $O(3)$ algebra constucted on unite vector
$p^1$. Not combersome manipulations lead to following explicit expression for d Alamber 
operator  
$$
(\cos \theta)^{-(\rho+1)}m^2(\cos \theta)^{(\rho+1)}=(\cos \theta)\partial_{\theta})^2+
\cot \theta \partial_{\theta}+(\cot \theta)^2\partial^2_{\phi}-(1+(\cos \theta)^2-
(\cos \theta)^2 l^2
$$

\subsection{ X-representation}

Terms $P,X$-representations we use in ussual sence for quantum mechanic. Propiar vallues 
of any 4 mutual commutative operators constructed from the elements of the algebra of 
quantized space may be used for construction of its basis. But basises such construction 
are not connected by point like transformations but only by canonical ones. To have the 
results in more nearest to ussual theory form it is suitable to consider basis connected 
with the propiar vallues of the following 4 operators
$$
x_4,\quad r^2\equiv (\vec x)^2-{(\vec l)^2\over M^2},\quad (\vec l)^2,\quad l_3
$$  
which can be interpreted physically as the time coordinate, direction in the space and 
distance up to point of obsevation from the inital point. We would not like to present 

\section{The $O(2,4)$ case $0\leq L^2$ connected with the $O(2,3)$ group of the motion}

$$
K_2=-\mu^2({I^2\over \mu^2}-{(x-{L^2\over H}p)^2\over \nu^2 L^2}+{L^2\over H^2}p^2+
{(f^2-l^2)\over H^2}),\quad 0\leq L^2,\mu^2=1-{H^2\over L^2M^2}
$$ 
In this case 7 generators
$$
l_{\alpha}=\epsilon _{\alpha,\beta,\gamma}Q_{\beta,\gamma},\quad p_4=T_{4,2}=P_{2,1}
\quad \bar x_{\alpha}=T_{\alpha,4}=Q_{\alpha,4}
$$
are compact one and 8
$$
I=T_{1,4}=\rho^1\cos \phi q^1_4+\rho^2\sin \phi q^2_4+\cos \phi \sum q^i_4 Q^{i,1}+
\sin \phi \sum_{2\leq i} q^i_4 Q^{i,2}+\sin \phi q^1_4 P^{2,1}
$$
$$
\bar x_4=T_{2,4}=-\rho^1\sin \phi q^1_4+\rho^2\cos \phi q^2_4-\sin \phi \sum q^i_4 
Q^{i,1}+\cos \phi \sum_{2\leq i} q^i_4 Q^{i,2}+\cos \phi q^1_4 P^{2,1}
$$
$$
p_{\alpha}=T_{1,{\alpha}}=\rho^1\cos \phi q^1_{\alpha}+\rho^2\sin \phi q^2_{\alpha}+
\cos \phi \sum q^i_{\alpha} Q^{i,1}+\sin \phi \sum_{2\leq i} q^i_{\alpha} Q^{i,2}+\sin 
\phi q^1_{\alpha} P^{2,1}
$$
$$
f_{\alpha}=T_{2,{\alpha}}=-\rho^1\sin \phi q^1_{\alpha}+\rho^2\cos \phi q^2_{\alpha}-
\sin \phi \sum q^i_{\alpha} Q^{i,1}+\cos \phi \sum_{2\leq i} q^i_{\alpha} Q^{i,2}+
\cos \phi q^1_{\alpha} P^{2,1}
$$
are noncompact. Now all notations with $Q,q$ are connected with four-dimensional 
compact group of rotation, $P,p$ with $O(2)$.   

\subsection{Invariant subspaces}

15 equations (\ref{BC}) are responcible for correct limit to the ussual theory. 
It is suitable to begin resolving of this system from 3 scalar equations
\be
(\vec l,\vec f)=0,\quad (\vec i,\vec p)=0,\quad (\vec l,\vec {\bar x})=0\label{AC}
\ee
Third equation means that the second Kazimir operator of four dimensional rotation 
group equal to zero. But Kazimir operators constructed from the generators of left or 
right reqular representation are the same. Thus we have
$$
(\vec l,\vec f)=Q^{12}Q^{34}+Q^{13}Q^{42}+Q^{14}Q^{23}=0
$$
Two second scalar equation after simple manipulations equivalent to
$$
\rho_1(\vec q^1\vec l)+\sum_{i=1}^4 (\vec q^i\vec l)Q^{i,1}=0,\quad
\rho_2(\vec q^2\vec l)+\sum_{i=2}^4 (\vec q^i\vec l)Q^{i,2}+(\vec q^1\vec l)P^{21}=0
$$
Further 
$$
(\vec q^i\vec l)=\sum \epsilon_{\alpha,\beta,\gamma}q^i_{\alpha}q^{\nu}_{\beta}q^{\mu}_
{\gamma}Q^{\nu,\mu}=\sum \epsilon_{i,\nu,\mu,t}q^t_4 Q^{\nu,\mu}
$$
Thus
$$
(\vec q^1\vec l)=\sum \epsilon_{1,\nu,\mu,t}q^t_4 Q^{\nu,\mu}
$$
which means that neither $\nu$ no $\mu$ not equal to 1 and $(\vec q^1\vec l)$ is the 
linear combination of the generators $Q^{2,3},Q^{2,4},Q^{3,4}$ with commuatation 
relaition of $O(3)$ algebra. By the same reasons
$$
(\vec q^2\vec l)=\sum \epsilon_{2,\nu,\mu,t}q^t_4 Q^{\nu,\mu}=aQ^{3,4}+bQ^{1,4}+cQ^{1,3}
$$
From this consideration it follows uniquely that under the additional condition
$$
\rho_2=0,\quad Q^{2,3}=0,\quad Q^{2,4}=0,\quad Q^{3,4}=0
$$
3 scalar equations (\ref{AC}) are sutisfied.
 
The last condition determines invariant subspace on which representation of $O(2,4)$ 
algebra (not of the general position $\rho_2=0$) is realised. These conditions means 
that representation of $O(4)$ algebra in its turn is realised only on one four-
dimensional unite vector $q^1_{\alpha}$ parametrised by 3 compact parameters. And the 
representation of $O(2,4)$ algebra of the considered type is realised on four compact 
parameters ( three parameters of unite four-dmesional $q^1$ vector and $\phi$. 

Now we substitute these results into the general formulae for non compact generators
$$
I=\cos \phi(\rho q^1_4+\sum q^i_4 Q^{i,1})+\sin \phi q^1_4 P^{2,1}=
\cos \phi A_4+\sin \phi B_4
$$
$$
\bar x_4=-\sin \phi (\rho q^1_4+\sum q^i_4 Q^{i,1})+\cos \phi q^1_4 P^{2,1}=
-sin \phi A_4+\cos \phi B_4
$$    
$$
p_{\alpha}=-\cos \phi(\rho q^1_{\alpha}+\sum q^i_{\alpha} Q^{i,1})-\sin \phi 
q^1_{\alpha} P^{2,1}=-\cos \phi A_{\alpha}-\sin \phi B_{\alpha}
$$
$$
f_{\alpha}=-\sin \phi(\rho q^1_{\alpha}+\sum q^i_{\alpha} Q^{i,1})+\cos \phi 
q^1_{\alpha}P^{2,1}=-\sin \phi A_{\alpha}+\cos \phi B_{\alpha}
$$
and would like to show that 12 remaining equations (\ref{BF}) are also satisfied.
As an example let us consider the first system of 3 equations. We have in a consequence
$$
If_{\alpha}+p_{\alpha}\bar x_4=(\cos \phi A_4+\sin \phi B_4)
(-\sin \phi A_{\alpha}+\cos \phi B_{\alpha})-(-\sin \phi A_4+\cos \phi B_4)
(\cos \phi A_{\alpha}+\sin \phi B_{\alpha})=
$$
$$
A_4 B_{\alpha}-B_4 A_{\alpha}-q^1_4 q^1_{\alpha}P^{12}=\sum (q^{\mu}_4 q^1_{\alpha}-
q^1_4 q^{\mu}_{\alpha}Q^{mu,1}=\sum q^{\mu}_4 q^{\nu}_{\alpha}Q^{\mu,\nu}P^{12}=
Q_{\alpha,4}P^{12}=p_4\bar x_{\alpha}
$$
All other equations (\ref{BC}) may be checked by the same way.

\subsection{Unitary representations}

In this case as it follows from explicit expression for $K_2$ the unitary 
representation of the main continues serie with $\rho=-2+i\sigma$ gives correct sign
in Kazimir operator to have  a correct limit $I\to 1$ to classical limit. By this 
reason we would not like to consider other unitary representations which are exist
in the case under consideration.

\subsection{de Alamber equation}

We will not present all generators of $O(4)$ in explicit form as functions of three
angles of unite $O(4)$ vector but only present below explicit form of the d Alamber
operator.  In case under consideration the group of motion is $O(2,3)$ and consequently
we have
$$
m^2=\vec p^2+\vec f^2-p_4^2-\vec l^2=(\cos \phi A_{\alpha}+\sin \phi B_{\alpha})^2+
(\sin \phi A_{\alpha}-\cos \phi B_{\alpha})^2-p_4^2-\vec l^2=
$$  
$$
\rho(\rho+3)-(\rho q^1_4+Q)^2+\partial^2_{\tau}+2\cot \tau \partial_{\tau}+\cot^2 
\tau\vec l^2-\cos^2 \tau \partial^2_{\phi}
$$
where $Q=\sum q^i_4 Q^{i,1}=-\sin  \tau \partial_{\tau},q^1_4=\cos \tau$. 
After a little further manipulations we come to a finally expression
$$
(\cos \tau)^{-\rho}m^2(\cos \tau)^{\rho}=(\cos \tau \partial_{\tau})^2+2\cot \tau 
\partial_{\tau}+\cot^2 \tau\vec l^2-\cos^2 \tau \partial^2_{\phi}
$$
As in the previous case ($O(3,3)$) variables in this equation are separated and without
any difficulties it is possible obtain the specter mass and corresponding wave 
functions. 

\section{The $O(2,4)$ case $L^2\leq 0$ connected with the $O(1,4)$ group of
motion}

As it follows from the explicit expression for second Kazimir operator
$$
K_2=-\mu^2({I^2\over \mu^2}-{(x-{L^2\over H}p)^2\over \nu^2 L^2}+{L^2\over H^2}p^2+
{(f^2-l^2)\over H^2}),\quad L^2 \leq 0\leq \mu^2=1-{H^2\over L^2M^2}
$$
is different from the previous one only by exchange $p\to \bar x,\bar x\to p,
f\to -f,l\to -l$. Thus corresponding formulae may be obtained by simple substitutions 
in corresponding formulae of the previous case.
 
7 generators
$$
l_{\alpha}=\epsilon _{\alpha,\beta,\gamma}Q_{\beta,\gamma},\quad \bar x_4=F_{4,2}=P_{2,1}
\quad p_{\alpha}=F_{\alpha,4}=Q_{\alpha,4}
$$
are compact one and 8
$$
I=F_{1,4}=\rho^1\cos \phi q^1_4+\rho^2\sin \phi q^2_4+\cos \phi \sum q^i_4 Q^{i,1}+
\sin \phi \sum_{2\leq i} q^i_4 Q^{i,2}+\sin \phi q^1_4 P^{2,1}
$$
$$
p_4=F_{2,4}=-\rho^1\sin \phi q^1_4+\rho^2\cos \phi q^2_4-\sin \phi \sum q^i_4 
Q^{i,1}+\cos \phi \sum_{2\leq i} q^i_4 Q^{i,2}+\cos \phi q^1_4 P^{2,1}
$$
$$
\bar x_{\alpha}=T_{1,{\alpha}}=\rho^1\cos \phi q^1_{\alpha}+\rho^2\sin \phi q^2_{\alpha}+
\cos \phi \sum q^i_{\alpha} Q^{i,1}+\sin \phi \sum_{2\leq i} q^i_{\alpha} Q^{i,2}+\sin 
\phi q^1_{\alpha} P^{2,1}
$$
$$
f_{\alpha}=F_{2,{\alpha}}=-\rho^1\sin \phi q^1_{\alpha}+\rho^2\cos \phi q^2_{\alpha}-
\sin \phi \sum q^i_{\alpha} Q^{i,1}+\cos \phi \sum_{2\leq i} q^i_{\alpha} Q^{i,2}+
\cos \phi q^1_{\alpha} P^{2,1}
$$
are noncompact. Now all notations with $Q,q$ are connected with four-dimensional 
compact group of rotation, $P,p$ with $O(2)$.

\subsection{Invariant subspaces and unitary representation}

It is not difficult to understand that the system of equations (\ref{BC}) is invariant
with respect to substitution $p\to \bar x,\bar x\to p,f\to -f,l\to -l$ and thus all
remains the same as in the previous section.              

\subsection{de Alamber operator}

The explicit expressions for quantized space generators are the following ones
$$
p_4=-\sin \phi(\rho^1q^1_4+ \sum q^i_4 Q^{i,1})+\cos \phi q^1_4 P^{2,1}
$$
$$
f_{\alpha}=-\sin \phi(\rho^1 q^1_{\alpha}+\sum q^i_{\alpha} Q^{i,1})+\cos \phi q^1_{\alpha} P^{2,1}
$$
$$
l_{\alpha}=\epsilon _{\alpha,\beta,\gamma}Q_{\beta,\gamma},\quad  p_{\alpha}=Q_{\alpha,4}
$$
The second Kazimir operator of $O(1,4)$ has the form:
$$
-m^2=p_4^2+f^2-p^2-l^2=\sum_{i=1}^4 (-\sin \phi(\rho q^1_i+\sum q^s_i Q^{s,1})+\cos \phi q^1_{\alpha} 
P^{2,1})^2-K_2(O(4))=
$$
$$
(\sin \phi \rho+\cos \phi \partial_{\phi})^2+3\cos \phi (\sin \phi \rho+\cos \phi \partial_{\phi})-
\cos \phi^2 K_2(O(4))
$$
Further regrouping of the terms lead to a final result
$$
-(\cos \tau)^{-(\rho+{3\over 2}}m^2(\cos \tau)^{\rho+{3\over 2}}=(\cos \phi \partial_{\phi})^2+
(l+{1\over 2})(l+{3\over 2})\cos \phi^2-{9\over 4}
$$

\section{The $O(1,5)$ case}
$$
K_2=\nu^2({p^2L^2\over H^2}+{x^2\over L^2\nu^2}+{f^2-l^2\over H^2}-{I^2\over \nu^2}),
\quad L^2\leq 0\leq \nu^2
$$
In this case 10 generators are compact 
$$
I=Q_{65},\quad p_{\alpha}=Q_{\alpha,5},\quad x_{\alpha}=Q_{\alpha,6},\quad l_{\alpha}=
\epsilon_{\alpha,\beta,\gamma}Q_{\beta,\gamma}
$$ 
(we use indexes  $(1,2,3,5,6)$ for numeration of all ingradientes of $O(5)$ algebra).
And 5 generators are non compact
$$
f_{\alpha}=\rho q^5_{\alpha}+\sum_{s=1}^5 q_{\alpha}^s Q^{(s,5)}\quad
x_4=\rho q^5_6+\sum_{s=1}^5 q_6^s Q^{(s,5)},\quad p_4=\rho q^5_5+\sum_{s=1}^5 q_5^s 
Q^{(s,5)}
$$

\subsection{Invariant subspaces}

Solution of the main system (\ref{BC}) is equivalent to equality to zero all generators 
of compact $O(4)$ algebra with generators $Q^{a,b}$ with $a,b=1,2,3,4$. This is equivalent
to realization of $O(5)$ algebra on unit vector $q^5_i$. This is not a big problem to check
this fact directly. 

\subsection{Unitary representation}

It is obvious that generators of the quantised space posses the unitary irreducible 
representation with $\rho=-2+i\sigma$. Compact generators are always antihermitian ones,
the condition of antihermitianes of noncompact generators leads directly to values for $\rho$ 
presented above. But this representation in contradiction with the classical limit $I\to 1$.
This easely to see from the explicit expressions for second Kazimir operator, which 
for this value of $\rho$ takes negative values. Thus as in the case of $O(3,3)$ algebra 
it is necessary to consider the case of unitary representation possess the Hermitian form.

The condition that invariance of hermitian form with respect to compact transformation 
restricted the functuonal dependence of the kernal of hermitian up to one variable
$K=K((q^5,\bar q^5))\equiv K(x)$. Equations responcible for invariance with respect of 
noncompact generators are the following ones
$$
[(\rho*+4)q^5_i+(\rho+4)\bar q^5_i-(\bar q^5_i-q^5_i x)\partial_x-(q^5_i-\bar q^5_i x)\partial_x]K=0
$$ 
which are equivalent to the pair of equations
$$
[(\rho*+4)+(\rho+4)x-(1-x^2)\partial_x]K=0,\quad [(\rho*+4)x+(\rho+4)-(1-x^2)\partial_x]K=0
$$
The condition of selfconsistenty leads to relation
$$
[\rho*-\rho](x-1)=0
$$ 
Condition $x=1$ leads to singular solition which responsible for the existence of representation 
with $\rho=-2+i\sigma$.

In the case of the real $\rho$ the kernal function has a form
$$   
K=c|1-x|^{-(\rho+4)}
$$
with all comments with respect to $O(3,3)$ section.

\subsection{d Alamber operator}

The group of motion in tyhis case is $O(1,4)$ and corresponding second order operator 
of Kazimir is the following one
$$
-m^2=\vec f^2+p_4^2-\vec p^2-\vec l^2=\sum_{a=1,2,3,4,}[(\rho+4) q^5_a+
\sum_{s=1}^5 Q^{(s,5)}q_a^s][\rho q^5_a+\sum_{s=1}^5 q_a^s Q^{(s,5)}]-\sum_{a\leq b}Q_{a,b}Q_{a,b}=
$$
$$
\rho(\rho+4)-[(\rho+4) \cos \tau-\sin \tau \partial_{\tau}][\rho\cos \tau-\sin \tau \partial_{\tau}]+
\sum_{i\leq j}Q_{i,j}Q_{i,j}-\sum_{a\leq b}Q_{a,b}Q_{a,b}=
$$
$$
\rho(\rho+4)-[(\rho+4) \cos \tau-\sin \tau \partial_{\tau}][\rho\cos \tau-\sin \tau \partial_{\tau}]+
\partial^2_{\tau}+3\cot \tau \partial_{\tau}-\cot^2 \tau l(l+2)
$$
Further regrouping of the terms leads to the finally expression
$$
(\cos \tau)^{-\rho}m^2(\cos \tau)^{\rho}=(\cos \tau \partial_{\tau})^2+
3\cot \tau \partial_{\tau}-(\cot \tau)^2l(l+2)
$$

\section{Outlook}

After this paper and \cite{I} are finished, the author hopes that 
his almost fourty years long struggle with quantized spaces \cite{3} 
is now over and our relations now will be more friendly. 

We have now a selfconsistent mathematical scheme, which permits
explicit calculations without making changes in the existing physical 
theory. The only small question that remains is whether the Nature
will accept the rules of the game we offer.

The new constants that arise in our machinery are of very large scale;
hence, if a confirmation of our constructions is to be found, it will 
come from cosmological observations. Unfortuntely, the author is very 
far from this field.

An essentially new feature of our formalism are the new terms in the 
energy balance (Lorenz moments). This gives a hope that the dark matter 
problem can be tackled from a new angle.

On the microscopic scale, the appearance of the de Sitter group 
essentially changes the problem of the unification of the internal and 
spatial symmetries. The non-invariance with respect to the time reversal
may have its consequences on the microscopic level as well.

\end{document}